\documentclass[%
reprint,
superscriptaddress,
 amsmath,amssymb,
floatfix,
]{revtex4-2}
\usepackage{lipsum} 

\usepackage[caption=false]{subfig}
\newcommand{\phantomsubfloat}[1]{
    {% apply caption setup only temporarily
        \captionsetup[subfigure]{labelformat=empty}
        \subfloat[][]{#1}
    }%
}
\usepackage{graphicx}% Include figure files
\usepackage{dcolumn}% Align table columns on decimal point
\usepackage{bm}% bold math
\usepackage{xcolor}
\usepackage[]{soul}
\DeclareMathOperator{\erf}{erf}

\begin{document}

\preprint{APS/123-QED}

\title{100 Gbps Integrated Quantum Random Number Generator Based on Vacuum Fluctuations}% Force line breaks with \\

\author{C\'{e}dric Bruynsteen}
\email{cedric.bruynsteen@imec.be}
\affiliation{%
 Ghent University-imec, IDLab, Dep. INTEC, 9052 Ghent, Belgium
}%

\author{Tobias Gehring}
\affiliation{
Center for Macroscopic Quantum States (bigQ), Department of Physics, Technical University of Denmark, 2800 Kongens Lyngby, Denmark
}%
\author{Cosmo Lupo}
\affiliation{
Dipartimento  Interateneo di Fisica, Politecnico \& Universit\`a di Bari, 70126, Bari, Italy
}%
\affiliation{
INFN, Sezione di Bari, 70126 Bari, Italy
}%

\author{Johan Bauwelinck}%
\author{Xin Yin}%

\affiliation{%
Ghent University-imec, IDLab, Dep. INTEC, 9052 Ghent, Belgium
}%

\date{\today}% It is always \today, today,

\begin{abstract}
Emerging communication and cryptography applications call for reliable, fast, unpredictable random number generators. Quantum random number generation allows for the creation of truly unpredictable numbers thanks to the inherent randomness available in quantum mechanics. A popular approach is using the quantum vacuum state to generate random numbers. While convenient, this approach was generally limited in speed compared to other schemes. Here, through custom co-design of opto-electronic integrated circuits and side-information reduction by digital filtering, we experimentally demonstrated an ultrafast generation rate of 100 Gbps, setting a new record for vacuum-based quantum random number generation by one order of magnitude. Furthermore, our experimental demonstrations are well supported by an upgraded device-dependent framework that is secure against both classical and quantum side-information and that also properly considers the non-linearity in the digitization process. This ultrafast secure random number generator in the chip-scale platform holds promise for next generation communication and cryptography applications.
\end{abstract}

\maketitle

%\tableofcontents
\section{Introduction} \label{sec:intro}
Random numbers are an essential resource in many applications such as cryptography \cite{Stipcevic2012QuantumCryptography}, statistical simulations \cite{Bauke2007RandomSimulations} or fundamental physical experiments. In cryptography the quality or randomness of the keys determines the security of the encryption, implying that truly unpredictable numbers are a crucial component of the modern digital society.
Pseudo-random numbers, while easy to generate, cannot be considered truly unpredictable due to their inherent deterministic behavior. As a result, physical phenomena have widely been adopted to generate truly random numbers. Quantum random number generators (QRNGs) harness the intrinsic randomness present in quantum mechanics to generate such numbers.

Numerous sources of entropy exist in quantum physics, with sources present in the field of photonics having shown to be very capable of conveniently generating random numbers at a high rate. Some examples of entropy sources are: photon number statistics \cite{ren_quantum_2011}, amplified spontaneous emission (ASE) \cite{williams_fast_2010,Martin2015}, vacuum noise \cite{Gabriel2010, Gehring2021}, laser phase noise \cite{jofre_true_2011} or Raman scattering \cite{collins_random_2015,england_efficient_2014}. The implementation complexity and attainable generation rate differs greatly depending on the source of entropy. Schemes that employ single-photon detectors (e.g.~photon number statistics) are generally limited in random number generation rate due to the low speeds of these detectors. Schemes making use of a continuous noise source can employ high-bandwidth photodiodes and can easily reach Gbps speeds \cite{zheng_6_2019,Avesani2018Source-device-independentGbps,Nie2015TheFluctuations,Bai202118.8Chip,Kordts2018SecurityGenerator}. However, not every scheme employing a continuous noise source can be easily integrated, making adoption outside lab environments more challenging. This is the case for noise sources like ASE and Raman scattering which can both achieve very high generation rates, but require Er/Yb-doped fibers \cite{williams_fast_2010,Martin2015} and non-linear crystals \cite{collins_random_2015,england_efficient_2014} respectively. Random number generation based on measuring phase noise requires either a laser driver circuit to generate narrow pulses \cite{jofre_true_2011, Abellan2016QuantumGeneration} or a feedback loop to create a very stable interferometer \cite{Nie2015TheFluctuations}, which both increase integration complexity. In this work vacuum noise will be used as the source of entropy, which is amplified by a balanced homodyne detector. This method of generating random numbers offers several advantages. First, the source of entropy, i.e. the vacuum noise, is readily available and therefore no bulky external components are required. A second advantage is the inherent cancelling of excess noise present in the Local Oscillator (LO) \cite{Gabriel2010} by using balanced detection, relaxing the requirements on the laser and increasing the resilience of the system against external perturbations.

The quality of the RNG is determined by its unpredictability. In order to obtain true unpredictability it is critical to map any imperfections in the measurement setup and to know how much information is available to the environment. Information is leaked to the environment via side-information channels, which can be classified as either classical or quantum. Classical side-information is any noise of a classical origin, e.g. noise generated by the electronics or relative intensity noise in the LO \cite{Haw2015MaximizationGenerator}. Quantum side-information channels arise due to the environment being entangled with the system used to extract the random numbers \cite{Frauchiger2013TrueDevices,Gehring2021}. Security proofs have been proposed which take into account classical side-information \cite{Haw2015MaximizationGenerator} as well as quantum side-information \cite{Gehring2021}. In addition to the side-information leakage, imperfect analog-to-digital converters (ADC) also give rise to reduced performance. Therefore the non-linear behavior of the ADC must also be accounted for \cite{Mitchell2015StrongGeneration, Gehring2021} to obtain an accurate estimate of the generation rate.

Besides accounting for the side-information leakage, much research effort has also been spent on increasing the speed of the random number generators. In general, the speed of vacuum noise based RNGs has been limited by the balanced homodyne detector speed and its noise performance. By integrating the homodyne detector, significant improvements have been demonstrated in the shot-noise limited bandwidth \cite{Tasker2021,Bruynsteen2021IntegratedMeasurements, Bai202118.8Chip}, enabling higher generation rates.

In this work we demonstrate a QRNG capable of delivering 100 Gbps of random numbers using an integrated balanced homodyne detector \cite{Bruynsteen2021IntegratedMeasurements}. This rate is achieved by employing a trusted, device-dependent security framework that takes into account both classical and quantum side-information channels and that is valid for any detector. In this framework, we first quantify this side-information for independent and identically distributed (i.i.d.) measurements. To extend the framework to non-i.i.d.~measurements, we build an effective model that maps the non-i.i.d.~case into an i.i.d.~one. To measure the static non-linearity of the ADC, a new method is provided and its impact on the generation rate is considered. Furthermore, the limited bandwidth of the receiver is augmented digitally by applying detector equalization. 
This improves the reliability of the effective i.i.d.~model to describe the non-i.i.d.~measurements and drastically boosts the generation rate.

\section{Min-entropy framework} \label{sec:min-entropy}
\begin{figure}[htbp]
    \centering
    \includegraphics[width = \linewidth]{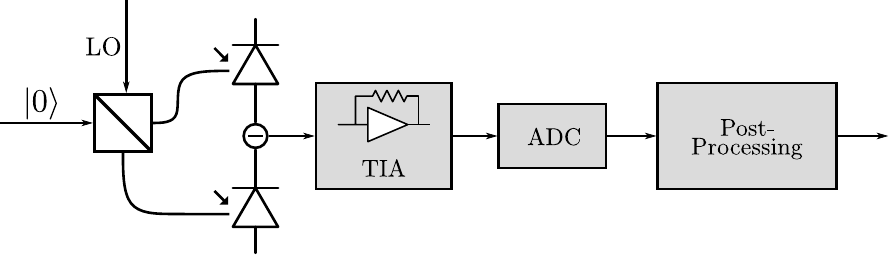}
    \caption{The block diagram for a quantum random number generator based on the quadrature measurement of the vacuum state $| 0 \rangle$.}
    \label{fig:QRNG_basic_overview}
\end{figure}

The QRNG generates random numbers by measuring an arbitrary quadrature \emph{Q} of a vacuum state $| 0 \rangle$ using homodyne detection. Practically, a balanced homodyne detector is used, consisting of an optical mixing element, a pair of balanced photodiodes and a transimpedance amplifier, which output is digitized using an ADC and further distilled to a sequence of true random bits. Figure \ref{fig:QRNG_basic_overview} shows a block diagram of a vacuum fluctuation based QRNG.

The probability density function of \emph{Q}, denoted by $p_Q$ is Gaussian with a mean of zero and variance $\sigma_Q^2$. In any practical implementation of a vacuum fluctuation based QRNG, the measured signal \emph{M} will not only consist of the useful signal \emph{Q}, but also contains traces of side-information. 
Two distinct side-information channels are considered here. The first form is denoted by \emph{E}, which contains the electronic and optical excess noise.
Furthermore, if the process is not i.i.d., a measurement output collected at a given time is correlated with past measured values, which yields a second side-information channel.

Previous work by Gehring et al. \cite{Gehring2021} has mapped the influence of these side-channels on the min-entropy under the assumption that the spectral shape of the vacuum noise and excess noise is identical. This implies that the quantum shot noise to classical excess noise clearance ratio, simply referred to as clearance, remains constant over the whole frequency range. In this work we extend this framework to be valid for any arbitrary spectral shape of the clearance.

\begin{figure*}[t]
    \includegraphics[]{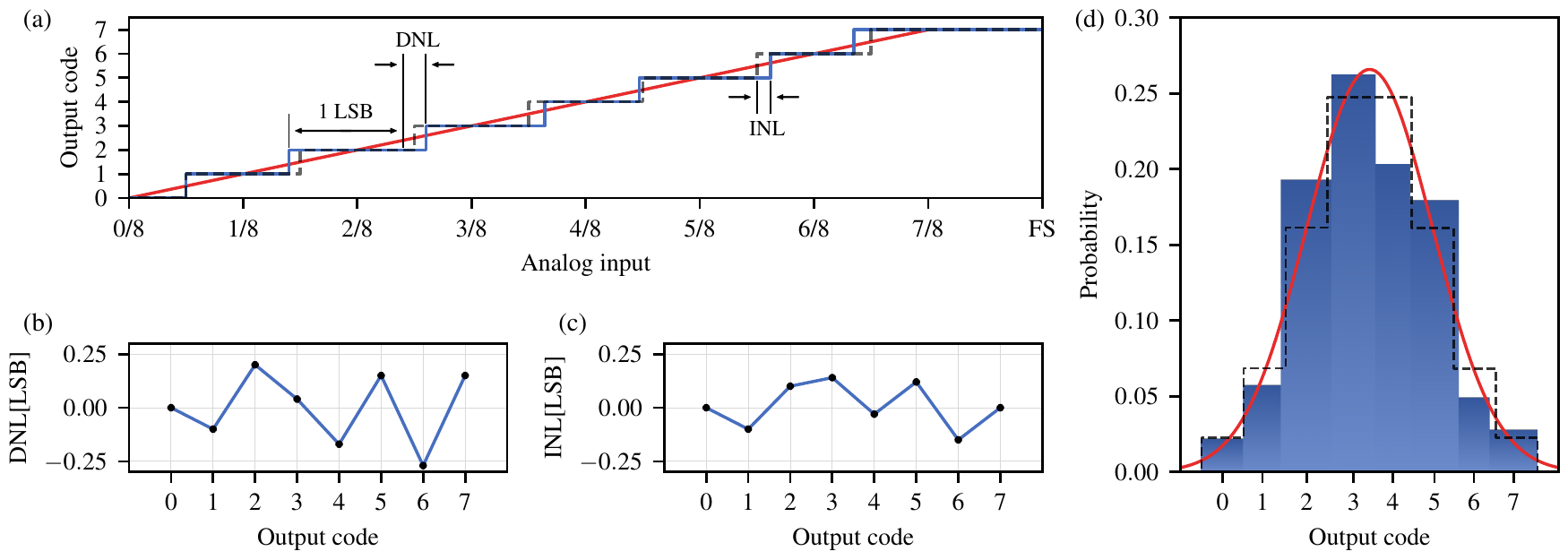}
    \phantomsubfloat{\label{fig:ADC_TF}}
    \phantomsubfloat{\label{fig:ADC_DNL_theory}}
    \phantomsubfloat{\label{fig:ADC_INL_theory}}
    \phantomsubfloat{\label{fig:ADC_Gaussian_theory}}
    \vspace{-2\baselineskip}
    \caption
     {\label{fig:theory_nonlinearity} \protect\subref{fig:ADC_TF}: Step response of a fictitious 3-bit ADC with non-linear behavior. The solid blue trace represents the actual response of the ADC and the dashed grey line represents the ideal step response. The solid red line is the analog input applied to the ADC. 
    \protect\subref{fig:ADC_DNL_theory}: The differential non-linearity (DNL) of the ADC. 
    \protect\subref{fig:ADC_INL_theory}: The integral non-linearity (INL) of the ADC.
    \protect\subref{fig:ADC_Gaussian_theory}: The output distribution of the 3-bit ADC assuming an input with a Gaussian density function is applied. The solid blue bars represent the actual response of the ADC, the dashed black line represents the response of an ideal 3-bit ADC and the solid red line represents the analog input.}
\end{figure*}

Let us first consider the i.i.d.~case, where only the first kind of side-information is present, due to excess noise.
We assume that the homodyne measurement, which contains traces of excess noise, can be described as a thermal state with mean photon number \emph{n} and gain factor \emph{g}. 
Let $M$ be the measured outcome, and $\sigma_M^2$ its variance.
When normalizing the vacuum noise variance to 1, the measurement variance is given by:
\begin{equation} \label{eq:sigma_mc2}
\sigma_{M}^2 = g^2(1+2n) \, ,
\end{equation}
where $\sigma_Q^2 = g^2$ is the variance of the vacuum fluctuations, and $\sigma_E^2 = 2 g^2 n$ is the variance of the excess noise.
Ideally, the mean photon number $n$ is small, which means that most of the noise contributing to the homodyne measurement is originating from the quantum shot noise.
Excess noise is therefore assumed to be Gaussian, and it is quantified by the parameter $n$.
In the worst-case scenario, the excess noise is due to entanglement between the optical mode that is measured and the environment. This indicates that our model is only valid for side-channels which are compatible with our Gaussian model and non-Gaussian entanglement is not accounted for.
To calculate the min-entropy for i.i.d.~measurements, a lower bound for i.i.d.~Gaussian states as described in \cite{Gehring2021} is used:
\begin{equation} \label{eq:hmin_eq}
H_\text{min} \geq
- \min_{\delta >0} \log{ \left[
\frac{(n+\delta)(1+n+\delta)}{\delta}
\, \mathcal{B} 
\right]} \, 
\end{equation}
\vbox{\noindent where $\delta$ is a parameter which is employed to improve the bound a posteriori and where:}
\begin{equation}
\begin{split}
\mathcal{B} & = \max\left\{ \text{erf}\left( \frac{\Delta x}{2 u} \right) , 
\frac{1}{2}\text{erfc}\left( \frac{R}{u} \right) 
\right\} \, , \\
u & = g \, \sqrt{ \frac{4n (n+1+\delta) + 2 \delta}{\delta} } \,.
\end{split}
\end{equation}
The min-entropy bound is determined by the ADC range \emph{R}, the ADC bin size $\Delta x$, and the gain $g$. Note that, as it should be expected, the min-entropy is a monotonically non-increasing function of $n$.
An optimum min-entropy is obtained when the gain $g$ is fixed in such a way that that the probability of triggering the edge bins are equal to the probability of triggering the bin corresponding to the center of the Gaussian distribution, resulting in: 
\begin{equation}
    \text{erf}\left( \frac{\Delta x}{2 u} \right) = \frac{1}{2}\text{erfc}\left( \frac{R}{u} \right)
\end{equation}
This results in an optimal lower bound for the min-entropy $H_{min}$:
\begin{equation} \label{eq:min_ent_optimized}
    H_{min} \geq -\log \left[\Gamma(n) \erf\left(\frac{\Delta x}{2u}\right)\right]
\end{equation}
where
\begin{equation}
   \Gamma(n) = (\sqrt{n}+\sqrt{n+1})^2
\end{equation}

An ideal N bit analog-to-digital converter (ADC) will have uniform bin sizes, equal to $\Delta x=R/2^N$ expressed in number of least significant bits (LSB). In reality the ADC will exhibit some non-linear behaviour, causing the bin sizes to vary over the output codes.
To map how much each individual bin size deviates from the ideal bin size, the differential non-linearity (DNL) \cite{Plassche2003} is employed. Codes which have a positive DNL give rise to a bin size larger than 1 LSB, and codes with a negative DNL to bin sizes smaller than 1 LSB. The accumulated DNL is represented by the integral non-linearity (INL) and is measured as the difference between the ideal ADC step response and the actual step response. Codes with larger bin sizes will get triggered more often compared to when an ideal ADC is used, resulting in a penalty in the available min-entropy. An example of a fictitious 3-bit ADC is shown in Figure \ref{fig:theory_nonlinearity}. As an upper bound for the min-entropy, the maximum DNL over all output codes is plugged in Equation \ref{eq:min_ent_optimized}:
\begin{equation} \label{eq:min_ent_ADC}
    H_{min} \geq -\log \left[\Gamma(n) \erf\left(\frac{R/2^N+ \text{DNL}_{max}}{2u}\right)\right]
\end{equation}

Due to the finite bandwidth of the detector, the measurement process can no longer be qualified as an i.i.d. stationary Gaussian process. The variance of the measurement \emph{M} can be written as $\sigma_M^2 = \sigma_{M,c}^2+\zeta$, with $\sigma_M^2$ the variance of the measured signal, $\sigma_{M,c}^2$ the variance of the measured signal conditioned on all the past measurements and $\zeta$ a factor which contains all the fluctuations of past measurements. We consider the quantum shot noise and excess noise to be statistically independent, leading to:
\begin{equation}
      \sigma_{M}^2= \sigma_{Q}^2+ \sigma_{E}^2 \;\; , \;\; S_M(f) = S_Q(f) + S_E(f)
 \end{equation}
where $S_M(f)$, $S_Q(f)$ and $S_E(f)$ are the power spectral densities (PSD) of the homodyne measurement, the quantum noise and excess noise respectively. 

We can also calculate the conditional variances for the homodyne measurement \emph{M}, the quantum signal \emph{Q} and excess noise \emph{E} based on the power spectral density \cite{Cover2006ElementsTheory}:
\begin{equation} 
\begin{split}
    \sigma_{Q,c}^2 &= \exp\left\{\int^{f_N}_0 \ln[f_NS_Q(f)]\frac{df}{f_N}\right\}, \\
    \sigma_{E,c}^2 &= \exp\left\{\int^{f_N}_0 \ln[f_NS_E(f)]\frac{df}{f_N}\right\}, \\
    \sigma_{M,c}^2 &= \exp\left\{\int^{f_N}_0 \ln[f_NS_M(f)]\frac{df}{f_N}\right\}
\end{split}
\end{equation}
where $f_N$ is the Nyquist frequency, which is equal to half the sampling rate.

The ratio between the conditional quantum variance and the conditional excess noise variance as well as the ratio between the conditional homodyne measurement and the conditional excess noise variance are related to the quantum shot noise to excess noise ratio called the clearance, denoted by $C(f)={S_Q(f)}/{S_E(f)}$. In general, $C(f)$ is a function of frequency. 
\begin{equation}\label{eq:ratios} 
\begin{split}
\frac{\sigma_{Q,c}^2}{\sigma_{E,c}^2} &= \exp\left\{\int^{f_N}_0 \ln\Big[{\frac{S_Q(f)}{S_E(f)}\Big]}\frac{df}{f_N}\right\} \\ 
 &= \exp\left\{\int^{f_N}_0 \ln[{{C}(f)]}\frac{df}{f_N}\right\} \\
 &= \mathcal{R}_c(C(f)) \\
\frac{\sigma_{M,c}^2}{\sigma_{E,c}^2} &= \exp\left\{\int^{f_N}_0 \ln\Big[{\frac{S_Q(f)+S_E(f)}{S_E(f)}\Big]}\frac{df}{f_N}\right\}\\
&= \mathcal{R}_c(C(f)+1)
\end{split}
\end{equation}

Here we define a function $\mathcal{R}_c(\cdot)$ for simplifying the expression of the conditional variances ratios, i.e., 
\begin{equation}
    \mathcal{R}_c(x(f)) = \exp\left\{\int^{f_N}_0 \ln[x(f)]\frac{df}{f_N}\right\}
\end{equation}
It is clear that, from the simplified expressions, the conditional variances ratios ${\sigma_{Q,c}^2}/{\sigma_{E,c}^2}$ and ${\sigma_{M,c}^2}/{\sigma_{E,c}^2}$ depend solely on the clearance $C(f)$. Also, both conditional variances ratios will increase when the clearance ratio becomes larger, as $\mathcal{R}_c(\cdot)$ is a monotonically increasing function for a real input. 

Using Eq. \ref{eq:ratios} and accounting for past measurements we obtain: 
\begin{equation} 
\sigma_{M}^2 = \sigma_{Q,c}^2+\sigma_{E,c}^2[\mathcal{R}_c(C(f)+1)-\mathcal{R}_c(C(f))]+\zeta
\label{eq:sigma_mc}
\end{equation}

In the particular case that the clearance is flat across the spectrum, Eq.~\ref{eq:sigma_mc} reduces to $\sigma_{M}^2 = \sigma_{Q,c}^2+\sigma_{E,c}^2+\zeta$ \cite{Gehring2021}.
We can identify the following from Eqs.~\ref{eq:sigma_mc2} and~\ref{eq:sigma_mc}:
\begin{equation} 
\begin{split}
    g^2&= \sigma_{Q,c}^2   \\
    2g^2n&=\sigma_{E,c}^2[\mathcal{R}_c(C(f)+1)-\mathcal{R}_c(C(f))] + \zeta
\end{split}
\end{equation}
In the worst-case scenario, this means that all correlations with past signals are due to entanglement with the environment, i.e. all correlations are quantum correlations.
From this we obtain the effective mean photon number \emph{n}:
\begin{equation}
\begin{split}
    n &= \frac{1}{2} \frac{\sigma_{E,c}^2[\mathcal{R}_c(C(f)+1)-\mathcal{R}_c(C(f))]+\zeta}{\sigma_{Q,c}^2} \\
      &= \frac{1}{2} \frac{\sigma_{M}^2}{\sigma_{Q,c}^2} - \frac{1}{2} \\
      &= \frac{1}{2} \frac{\sigma_{M}^2}{\sigma_{M,c}^2}\mathcal{R}_c\left(1+\frac{1}{C(f)}\right) - \frac{1}{2}
\end{split}
\label{eq:mean_phot_nr}
\end{equation}

Equation \ref{eq:mean_phot_nr} is an effective i.i.d.~model to describe the non-i.i.d.~measurements.
This allows us to use Eq.~\ref{eq:min_ent_ADC} to estimate the min-entropy of the non-i.i.d.~case.
Equation \ref{eq:mean_phot_nr} also shows that the effective mean photon number scales proportionally to the temporal correlations (${\sigma_{M}^2}/{\sigma_{M,c}^2}$) present in the homodyne measurement, as well as inversely to the clearance. The temporal correlations can be improved by making sure that the frequency response of the homodyne measurement remains constant, in which case the temporal correlations become equal to 1. Interestingly, since the mean photon number is dependent on $1+1/C(f)$, it is not necessary for the clearance to have a flat frequency response. Indeed, as long as the clearance is sufficiently large, the factor $1+1/C(f)$ will be close to 1. The mean photon number hence indicates that an ideal detector exhibits a flat frequency response and a high clearance level over a wide frequency range, limiting the amount of both quantum and classical side-information.

\begin{figure*}[htbp]
\centering
\subfloat{\label{fig:Hmin_vs_tempcorr}}
{\includegraphics[scale =0.9]{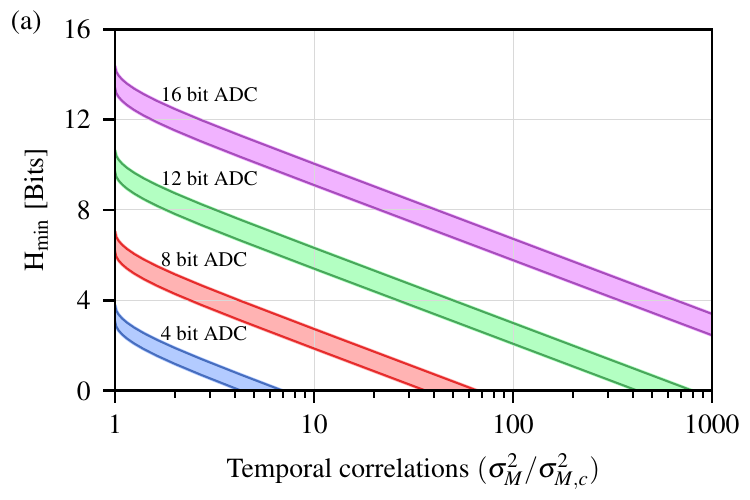}}
\subfloat{\label{fig:Hmin_vs_CLR}}
{\includegraphics[scale =0.9]{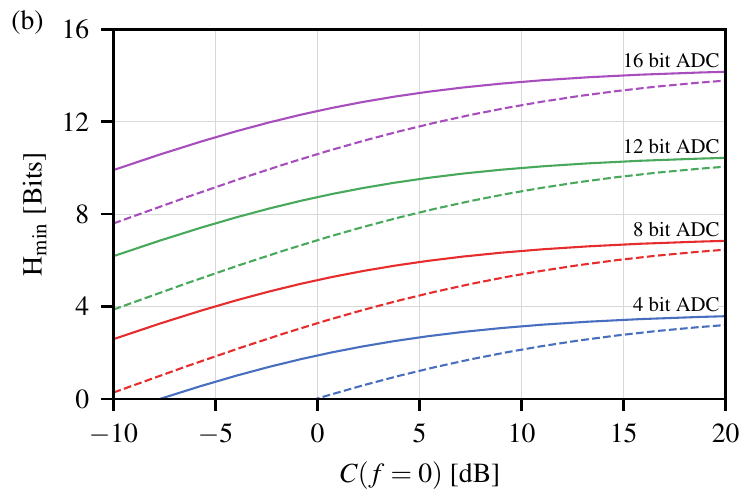}}
\caption{\protect\subref{fig:Hmin_vs_tempcorr}: Min-entropy for 4, 8, 12, and 16 bit ADC resolution versus the temporal correlation factor ${\sigma_{M}^2}/{\sigma_{M,c}^2}$.
Here $\sigma_{M,c}^2$ and $\sigma_{M}^2$ are the conditional and unconditional variance of the homodyne measurement respectively. The shaded areas indicate the regions for the ADC linearity varying between a bin width $\Delta x = 1$ LSB, the upper trace, and a bin width $\Delta x = 2$ LSB, the lower trace. \protect\subref{fig:Hmin_vs_CLR}:  Min-entropy for 4, 8, 12, and 16 bit ADC resolution versus the DC clearance level. The solid traces represent the scenario where the clearance is constant over the complete frequency range. The dashed traces represent the scenario where the clearance follows a second order Butterworth response with a bandwidth of $\frac{f_N}{5}$. The assumption is made here that the ADC is ideal and the temporal correlations are equal to one.}
\end{figure*}

The min-entropy for different temporal correlations (${\sigma_{M}^2}/{\sigma_{M,c}^2}$), ADC resolutions and bin width assuming a noiseless receiver, i.e. $\mathcal{R}_c(1+\frac{1}{C(f)}) =1$, is plotted in Figure \ref{fig:Hmin_vs_tempcorr}. The upper trace represents the min-entropy when the bin width is equal to 1 LSB, the bottom trace when the bin width is equal to 2 LSB ($\text{DNL}_{max}=1$). As expected, when the temporal correlations become larger the min-entropy reduces, as well as when the ADC reduces in linearity. Besides the temporal correlations the clearance also is a major factor in determining the QRNG performance. The solid traces in Figure \ref{fig:Hmin_vs_CLR} show the min-entropy for when the clearance $C(f)$ remains constant over frequency but with a varying amplitude. It is clear from the figure that it is not a requirement to have a clearance with amplitude larger than 0 dB to achieve a non-zero min-entropy. However, the assumption of a constant clearance over frequency is not guaranteed for each detector. This assumption is typically no longer valid for high bandwidth detectors making use of optimized transimpedance amplifiers. The shape of the frequency response of the clearance can be reasonably well approximated by the inverse of the input referred current noise density of the transimpedance amplifier. At high frequencies the input referred current noise density scales proportional to $f^2$ \cite{sackinger_analysis_2017, carusone2011analog}. This indicates that at high frequencies the clearance decreases proportional to $1/f^2$, which has indeed been experimentally verified \cite{Bruynsteen2021IntegratedMeasurements}. The dotted lines in Figure \ref{fig:Hmin_vs_CLR} represent the situation of the clearance following a second order Butterworth response with bandwidth $f_{BW}=\frac{f_N}{5}$. The nonuniform spectral response of the clearance has an adverse effect on the min-entropy, which was not captured in previous security proofs \cite{Haw2015MaximizationGenerator, Gehring2021}.

\section{Detector equalization} \label{sec:channel_eq}

From the previous sections we have concluded that a high bandwidth receiver with low temporal correlations is required to realize a high performing QRNG. The receiver, which acts as a low-pass filter, has a different group delay in its passband compared to its high frequency stopband. This causes temporal correlations to be present in the output signal. If the bandwidth of this filter is lower than the ADC's Nyquist frequency, then the temporal correlations will also manifests themselves at the output of the ADC. Unfortunately, it is increasingly difficult to design a low-noise transimpedance amplifier that exhibits a large bandwidth ($>$10 GHz) while also maintaining a high degree of sensitivity ($>$10 dB clearance), due to an intrinsic trade-off between noise and bandwidth \cite{sackinger_transimpedance_2010}. This is problematic for achieving a high generation rate, which requires both a flat frequency response and a sufficiently large clearance (Section \ref{sec:min-entropy}). To solve the problem of temporal correlations, usually referred to as intersymbol interference (ISI) in traditional telecom applications, often equalizers are employed. When the channel is subjected to linear distortions, an equalizer can apply the inverse transfer function of the channel, effectively cancelling any temporal correlations. This type of equalizer typically is referred to as a zero-forcing equalizer, which is an ideal equalizer for only cancelling ISI \cite{Proakis2013FundamentalsSystems}. The output homodyne detector is linearly proportional to the vacuum noise (with gain factor \emph{g}), making it ideal for equalization. The outcome after applying the equalization filter is a flattened spectral response with significant reduced temporal correlations \cite{Kordts2018SecurityGenerator}.

\begin{figure}[h!]
    \centering
    \includegraphics[width = \linewidth]{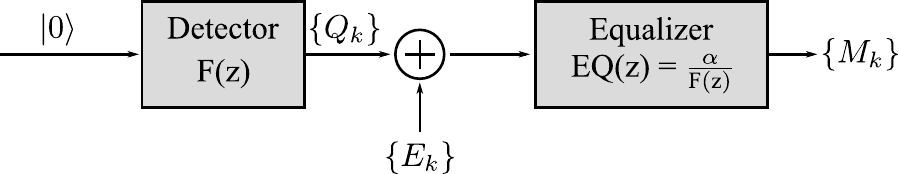}
    \caption{The block diagram for a zero-forcing linear equalizer \cite{Proakis2007DigitalEdition}.  $| 0 \rangle$ represents the input shot noise, $Q_k$ represents the shot noise sampled after passing through the detector at time k, $E_k$ represents the excess noise and $M_k$ represents the outcome of the homodyne measurement. }
    \label{fig:channel_eq}
\end{figure}

Equalization can be applied either in the analog domain via the use of an analog filter (e.g. a continuous time linear equalizer \cite{Peng2017ACMOS}) or can be accomplished in the digital domain after digitization. In this work we have chosen to apply a digital equalizer, which is easily reconfigurable and has the ability to compensate for very steep frequency responses. A block diagram of the detector with a zero-forcing linear equalizer is shown in Figure \ref{fig:channel_eq}. The digital frequency response of the detector F(z) is influenced by different frequency limiting factors, such as the opto-electrical bandwidth of the photodiodes, the bandwidth of the transimpedance amplifier and the bandwidth of the anti-aliasing filter.

EQ(z) can be implemented as a finite input response (FIR) filter which is equal to the inverse of the detector response F(z) scaled with a factor $\alpha$. The shape of this equalization filter, determined by F(z), takes care of removing the ISI. The DC gain factor $\alpha$ guarantees that the energy of the signal does not increase, which means that the variance of the measured signal does not change by adding the equalization filter, i.e. $\sigma_{M,\mathrm{no \ eq}}^2 = \sigma_{M,\mathrm{eq}}^2$. The amount of coefficients necessary to implement the FIR filter depends on the detector response F(z). If F(z) demonstrates abrupt changes in its spectral shape, a large amount of FIR taps are required to efficiently remove ISI. The values of the FIR tap coefficients are determined by performing a least-square fit on the inverse of F(z). Finally, the FIR coefficients are scaled with a factor $\alpha$ to establish the DC gain.

\section{Practical implementation} \label{sec:experimental}
\begin{figure*}[h]
    \includegraphics[width = \linewidth]{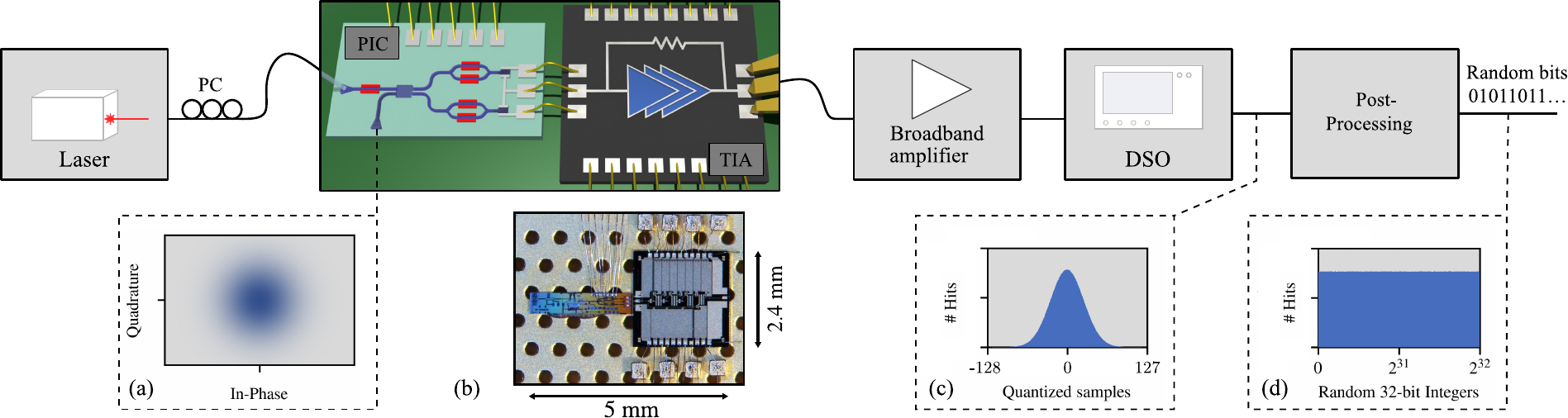}
    \caption{\label{fig:QRNG_system_overview} An overview of the QRNG setup. Subfigure (a) shows the vacuum noise which is used as a source to generate random numbers. Subfigure (b) shows a micrograph of the manufactured PIC and TIA. Subfigure (c) shows the Gaussian distribution after digitization. Subfigure (d) shows the distribution of distilled random 32-bit integers, grouped into 256 bins.}
    \end{figure*}

The practical implementation of the QRNG is shown in Figure \ref{fig:QRNG_system_overview}. A 1550nm CW laser (Koheras Basik E15) is fed into a photonic integrated circuit (PIC) fabricated using imec’s iSiPP50G silicon photonics platform. The PIC contains a tunable 2x2 mixing element connected to two photodiodes. The photocurrent is converted to a voltage by a custom transimpedance amplifier fabricated in a 100nm GaAs pHEMT technology and amplified by a linear broadband amplifier (SHF 807) to optimally fill the ADC range. Next, the analog signal is digitized by a Keysight DSOZ632A digital storage oscilloscope (DSO) which has an internal 8 bit ADC at a sample rate of 20 GS/s. The captured data is processed offline to equalize the detector response and to generate random bitstreams based on the proposed min-entropy framework.

Besides the obvious reduction in size, the use of a custom integrated balanced homodyne detector offers the possibility to design circuits which operate optimally for the application at hand, and therefore greatly improve performance compared to discrete, off-the-shelf implementations. This is reflected in the large shot-noise limited bandwidth of 20 GHz and the high maximum clearance of 28 dB \cite{Bruynsteen2021IntegratedMeasurements}. Furthermore, the frequency response of the detector has a gradual gain roll-off at high frequencies, limiting the amount of FIR taps required to equalize the detector response. Compared to QRNGs which use discrete components or integrated off-the-shelf components for either the optical front-end or the TIA \cite{Gehring2021, Haw2015MaximizationGenerator, Honz2021BroadbandGeneration, Bai202118.8Chip, Kordts2018SecurityGenerator}, the achievable speed and sensitivity are greatly improved thanks to the co-design between the PIC and TIA, the significantly smaller packaging parasitics and to the use of small, high-speed integrated photodiodes.

\begin{figure*}[!ht]
\centering
\begin{minipage}{\linewidth}
\subfloat{\label{fig:heatmap_equalise_extended}}{\includegraphics{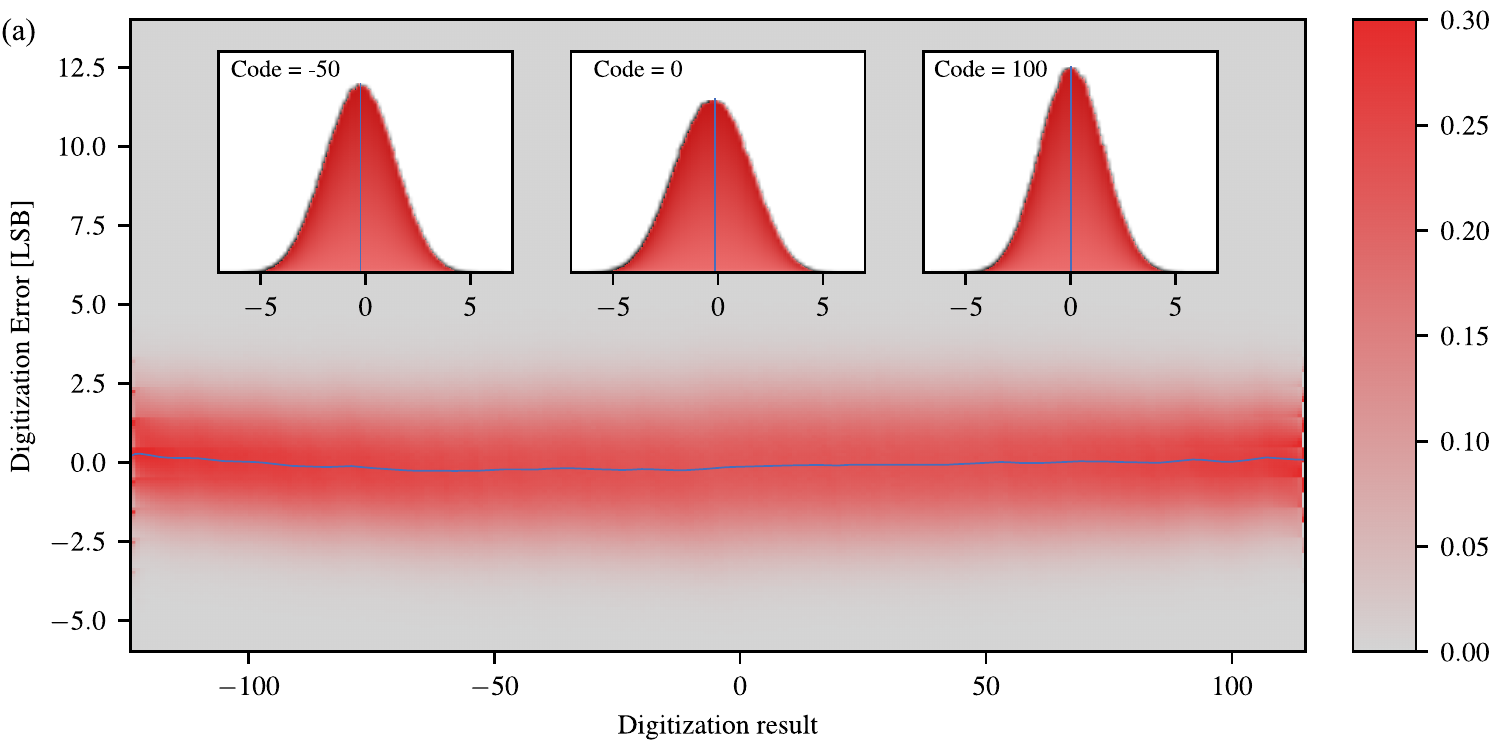}}
\end{minipage}

\begin{minipage}{0.49\linewidth}
\subfloat{\label{fig:INL}}{\includegraphics{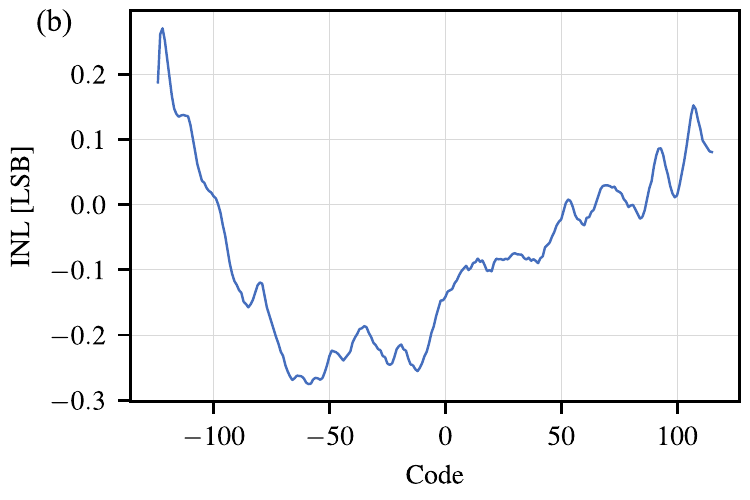}}
\end{minipage}
\begin{minipage}{0.49\linewidth}
\subfloat{\label{fig:DNL}}{\includegraphics{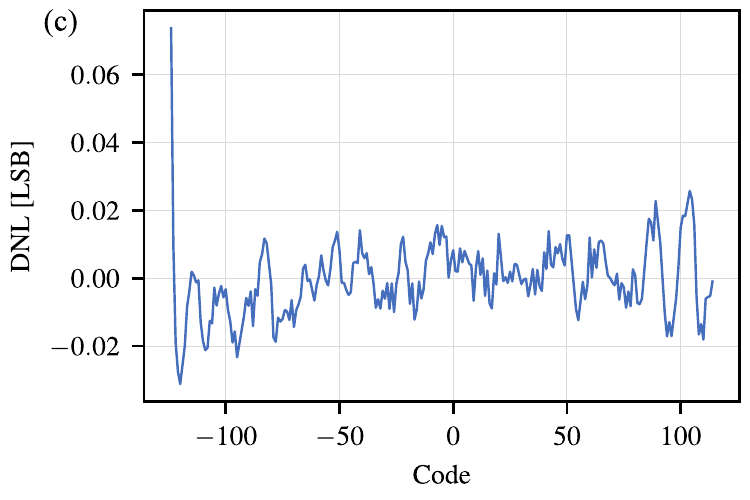}}
\end{minipage}
\caption{\protect\subref{fig:heatmap_equalise_extended} Heatmap of the errors present in the captured sine wave. Inset are the distributions for the codes -50, 0 and 100. The blue line represents INL, i.e. the mean of the errors for each digitization result.
\protect\subref{fig:INL} The integral non-linearity versus the ADC code. \protect\subref{fig:DNL} The differential non-linearity versus the ADC code. } 
\end{figure*}

Using both the min-entropy framework of Section \ref{sec:min-entropy} and the detector equalization of Section \ref{sec:channel_eq}, it is now possible to measure the metrics which are required to estimate the generation rate for the setup shown in Figure \ref{fig:QRNG_system_overview}. The various metrics which should be obtained are: the maximum DNL of the ADC, the PSD of the homodyne measurement and the excess noise, and finally the FIR filter tap coefficients.
\subsection{ADC non-linearity}

To measure the DNL, a method based on the IEEE Std 1241-2010 \cite{Tilden2011IEEEStd.1241-2000} is used. We apply a sine wave generated by an RF signal generator to the input of the oscilloscope. The frequency of the sine was chosen such that the sampling frequency is a non-integer multiple of the sine wave frequency. This guarantees that each time the sine wave is sampled, a different phase is measured, preventing only a select few phases being captured. 

\noindent This implies:
\begin{equation}
    f_{sine} = f_s\left(\frac{J}{M}\right) 
\end{equation}
Where $f_{sine}$ is the frequency of the input sine wave, $f_s$ is the sampling frequency, $M$ is the record length and $J$ is relatively prime to $M$. An alternative to the sine wave would be to generate a sawtooth or a triangle wave \cite{Mitchell2015StrongGeneration} using an arbitrary waveform generator. However, it is much more straightforward to generate a pure sine tone instead of more spectrally intensive waveforms such as a sawtooth or a triangle wave. If these would be used, it is critical that the digital-to-analog converter generating these signals has a known non-linearity, or is much more linear than the analog-to-digital converter under test.

The amplitude of the sine wave is chosen such that the full range of the ADC is being utilized without excessive clipping. At this point, the captured data deviates from an ideal sine wave due to the addition of non-linear behavior and noise. The noise captured in this measurement is not representative for the excess noise present in the setup of Figure \ref{fig:QRNG_system_overview}, and hence should be separated from the non-linearity. To this end, a sine wave is fitted to the captured data using the following form:
\begin{equation}
    x[n] = A\sin({2\pi f_0t_n+ \phi_0})+B
\end{equation}
In the above form, \emph{A} is the amplitude, $f_0$ is the frequency, $\phi_0$ is the phase and \emph{B} is the offset. A least square optimization is used to obtain these fitted parameters. After fitting, the error between the measured data and the ideal fitted curve can be calculated for each code of the ADC.

A 125.0125 kHz sine wave, generated using an Agilent N5182A MXG Vector Signal Generator, was applied as input to the oscilloscope. The oscilloscope captured sufficient periods of the sine wave such that each code is triggered more than one million times. This resulted in 400 MSamples being captured using the oscilloscope's 8 bit ADC. 

The digitization error for each code is plotted in Figure \ref{fig:heatmap_equalise_extended}. Each point along the x-axis contains a deterministic part, i.e. the non-linearity, and a stochastic part, i.e. noise. The distribution is shown for the codes -50, 0 and 100 (Fig. \ref{fig:heatmap_equalise_extended}), and can be described by a Gaussian distribution. The non-linear behavior of the ADC manifests itself in the mean of this distribution, and is equal to the integral non-linearity (INL) \cite{Plassche2003} of the ADC. The INL is marked by a blue line on Figure \ref{fig:heatmap_equalise_extended}. A zoomed version of the INL is shown in Figure \ref{fig:INL}. As the INL is simply the accumulated DNL \cite{Plassche2003}, a straightforward conversion between INL and DNL is obtained by calculating the difference in INL between each successive code (Fig. \ref{fig:DNL}). The maximum DNL is 0.074 LSB, which means that the maximum bin size will be 1.074 LSB.

\begin{figure}[htbp]
    \centering
    \begin{minipage}{\linewidth}
    \subfloat[No equalization.]{\includegraphics[]{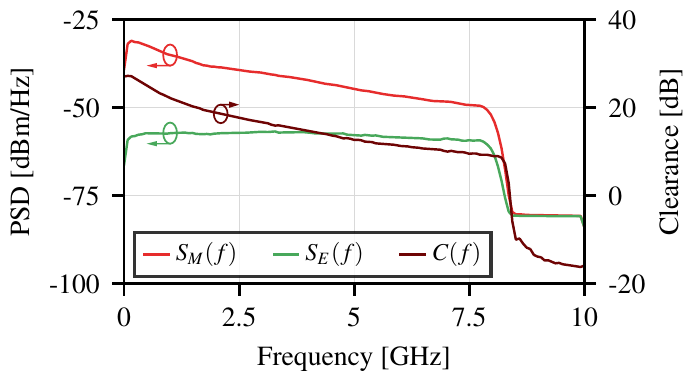}\label{fig:spectrum_before_eq}}
    \end{minipage}
    \begin{minipage}{\linewidth}
    \subfloat[Equalization up to 8 GHz.]{\includegraphics[]{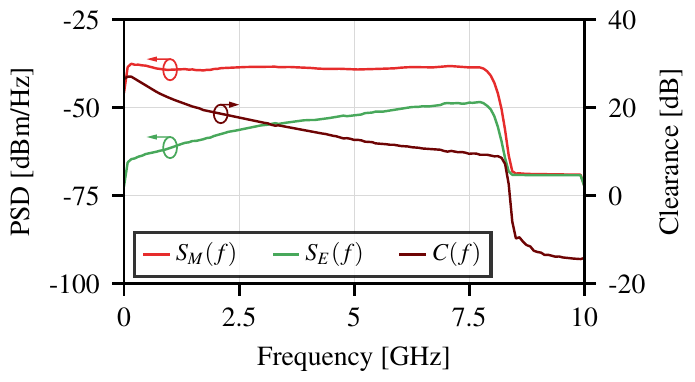}\label{fig:spectrum_8Ghz_eq}}
    \end{minipage}
    \begin{minipage}{\linewidth}
    \subfloat[Equalization up to 10 GHz.]{\includegraphics[]{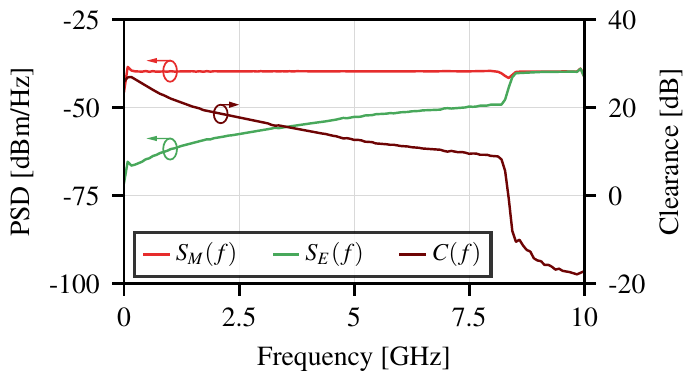}\label{fig:spectrum_10Ghz_eq}}
    \end{minipage}
    \caption{Subfigures \protect\subref{fig:spectrum_before_eq}, \protect\subref{fig:spectrum_8Ghz_eq} and \protect\subref{fig:spectrum_10Ghz_eq} display the homodyne measurement PSD $S_M(f)$, the excess noise PSD $S_E(f)$ and the clearance $C(f)$ for the cases where respectively no equalizer, an equalizer up to 8 GHz and an equalizer up to 10 GHz is used.}\label{fig:spectrum}
\end{figure}

\subsection{PSD and detector equalization}
In order to obtain the clearance $C(f)$, the FIR equalization filter and the temporal correlations ${\sigma_{M,c}^2}/{\sigma_{M}^2}$, the PSD of the homodyne measurement and the excess noise must be determined. The PSD of the homodyne measurement is approximated by running the setup shown in Figure \ref{fig:QRNG_system_overview} while the PSD of the digitized output is estimated using Welch's method \cite{Welch1967ThePeriodograms}. The balanced homodyne detector has been demonstrated to operate in a shot-noise limited regime \cite{Bruynsteen2021IntegratedMeasurements}, therefore the excess noise PSD is estimated by also running the setup shown in Figure \ref{fig:QRNG_system_overview}, but now with the laser turned off. Figure \ref{fig:spectrum_before_eq} shows the two densities as well as the ratio of both densities, i.e. the clearance. The clearance drops to very low values ($<$-10 dB) in the frequency band 8 GHz - 10 GHz. This drop in clearance originates from the internal anti-aliasing filter of the oscilloscope, causing the vacuum noise to be suppressed and be dominated by excess noise in this frequency band. This filter has a cutoff frequency of 8 GHz when the sampling rate is set to 20 GSamples/s. The temporal correlation factor before applying the equalizer is 8.388.

Because the roll-off of the anti-aliasing filter is steep, a high order FIR filter is required to adequately compensate for this frequency response. Alternatively it is also possible to only compensate the frequency response up to 8 GHz. Because the roll-off from 0 Hz to 8 GHz is relatively moderate, only a limited amount of filter taps are required. Both options are implemented. Figure \ref{fig:spectrum_8Ghz_eq} shows the power spectral densities and clearance which have been equalized up to 8 GHz. The equalization filter in this scenario uses 9 FIR filter taps. When comparing Figure \ref{fig:spectrum_before_eq} and Figure \ref{fig:spectrum_8Ghz_eq}, it is apparent that the clearance is identical for both scenarios because the filter is applied to the vacuum noise and the excess noise simultaneously. Because the original transfer function follows a lowpass characteristic, the 8 GHz equalizer attenuates low frequency components and boosts high frequency components. The result of applying this equalizer is that the temporal correlation factor has improved to 2.79. 

\begin{figure}[h]
    \centering
    \includegraphics[]{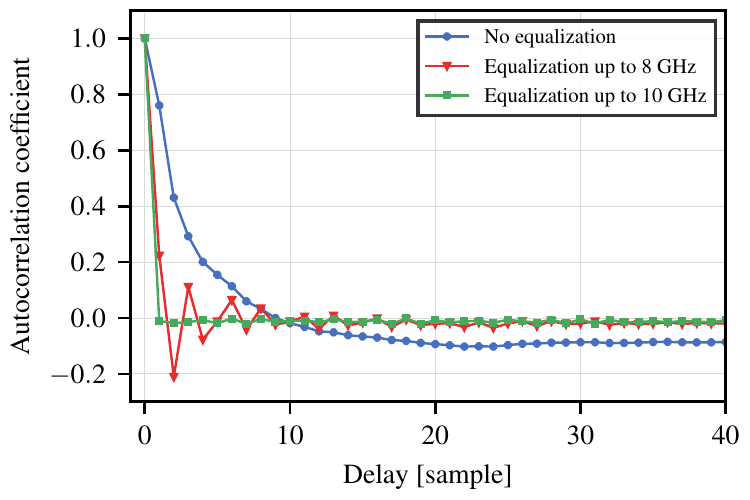}
    \caption{The effect of the equalizers on the autocorrelation. The autocorrelation coefficients are averaged over 10,000 measured datasets.}
    \label{fig:correlations_homodyne}
\end{figure}

Finally a filter is fitted to the complete 10 GHz spectrum (Fig. \ref{fig:spectrum_10Ghz_eq}). The filter order was chosen to be 201, which can be reasonably implemented in real-time on an FPGA or ASIC \cite{DeDinechin2010AFPGA,Bae2019ImprovedFilters}. Removing the bandwidth limitations added by the anti-aliasing filter further reduces the temporal correlation factor to 1.004. To verify the improvement in temporal correlation, an equalization filter was fitted to one set of measurement data and applied to another, independent, set of data. When examining the autocorrelation of the second set of data before and after equalization (Fig. \ref{fig:correlations_homodyne}), it becomes clear that the temporal correlations indeed have improved significantly after equalization.

\subsection{Generation rate}
Combining the results from the previous subsections, the min-entropy can be calculated. The generation rate is obtained using the leftover hash lemma \cite{Tomamichel2011LeftoverInformation} with a security factor $\epsilon_{hash}=10^{-14}$ and input word size n = 8192 bits.

Before applying any equalization the high temporal correlations factor of 8.388 limits the min-entropy $H_{min}$ to only 2.01 bits, which results in a generation rate of 38.13 Gbps. Applying a 9-tap equalizer up to 8 GHz, the min-entropy increases to 3.70 bits, yielding a generation rate of 71.88 Gbps. Ultimately, when applying a 201-tap full spectrum equalizer the min-entropy increases to 5.09 bits and, thanks to the further improvements in temporal correlations, the generation rate is increased to 100 Gbps. Theoretically, an ideal detector which at the same time exhibits no temporal correlations, an infinite clearance and no ADC non-linearity can achieve a min-entropy of 7.04 bits and a generation rate of 138.75 Gbps. Comparing the results of our detector to the ideal detector, it is apparent that  a lot of performance is lost when the temporal correlations are high. However, by applying detector equalization we are able to significantly boost the generation rate and approximate to the theoretic rate obtained by an ideal detector.

We remark that the equalization does not increase the min-entropy of the source, as the latter is due to quantum fluctuations and cannot be increased by post-processing. 
The role of the equalization is to improve the effective i.i.d.~model that we use to compute the min-entropy rate of the non-i.i.d.~measurements.
In fact, Eq.~\ref{eq:mean_phot_nr} shows that the effective mean photon number can be minimized by reducing the ratio 
$\sigma_{M}^2 / \sigma_{M,c}^2$, whereas the clearance $C(f)$ remains unaffected by the equalization.

Random numbers were extracted using a Toeplitz hashing algorithm with matrix dimensions n = 8192 bits and m = 5120 bits for the 100 Gbps case and were applied to both the dieharder \cite{dieharder} and NIST \cite{NIST2010} statistical batteries of tests. The generated random numbers passed both sets of tests.

\section{Conclusion}
In this work an integrated quantum random number generator based on vacuum fluctuations achieving a 100 Gbps generation rate is demonstrated. This generation rate is obtained by applying a framework secure against both classical and quantum side-information. A method for measuring the static ADC non-linearity is established. Next, by applying detector equalization the finite bandwidth present in the receiver is compensated, which reduces the amount of penalty induced by quantum side-information leakage. The achieved rate of 100 Gbps is significantly faster compared to other recent random generators based on vacuum fluctuations \cite{Gehring2021, Haw2015MaximizationGenerator, Honz2021BroadbandGeneration, Bai202118.8Chip, Kordts2018SecurityGenerator}. Previously, one of the limiting factors to achieve a high generation rate using vacuum fluctuations was the presence of classical noise in the measurement \cite{Herrero-Collantes2017QuantumGenerators}. By using custom, application-specific, integrated circuits we demonstrate that this bottleneck can be greatly reduced, proving that a vacuum fluctuations based QRNG is a viable solution for applications demanding high generation rates.

\section{Acknowledgements}
This work was supported by the Research Foundation Flanders through FWO Weave project SQOPE (G092922N) and the Quantum-flagship UNIQORN project that has received funding from the EU Horizon 2020 Research and Innovation Programme under grant agreement No. 820474. C.B. acknowledges support from the Research Foundation Flanders through FWO fellowship 1SB1721N. T.G. acknowledges support by the Danish National Research Foundation, Center for Macroscopic Quantum States (bigQ, DNRF142).
C.L. acknowledges support from the European Union – Next Generation EU: NRRP Initiative, Mission 4, Component 2, Investment 1.3 – Partnerships extended to universities, research centres, companies and research D.D. MUR n. 341 del 15.03.2022 – PE0000023 – ``National Quantum Science and Technology Institute".

\section{Author contributions}
C.B. designed the integrated devices, obtained the main experimental results and wrote the manuscript draft. C.B., T.G. and X.Y. performed data analysis. T.G., C.L., C.B. and X.Y. contributed to the theoretical framework. X.Y. and J.B. supervised the project.
All authors contributed to the interpretation, discussion of the results and writing of the manuscript.

\section{Data availability}
Data that support the plots within this paper and other findings of this study are available from the corresponding authors upon reasonable request.
\section{Competing interests}
The authors declare no competing interests.

\bibliography{references.bib}% Produces the bibliography via BibTeX.

\end{document}